\documentclass[singlecolumn]{jpsj3}
\usepackage{graphicx}
\usepackage{txfonts}
\usepackage[usenames]{color}
\title{Borel-Pad\'e re-summation of the $\beta$-functions describing Anderson localisation in the Wigner-Dyson symmetry classes}

\author{\name{Yoshiki \surname{Ueoka}}\thanks{Present address: Division of Electrical, Electronic and Information Engineering, Graduate School of Engineering, Osaka University, 2-1 Yamadaoka, Suita, Osaka 565-0871, Japan}, \name{Keith \surname{Slevin}}}
\inst{Department of Physics, Graduate School of Science, Osaka University, 1-1 Machikaneyama, Toyonaka, Osaka 560-0043, Japan}

\abst{We describe a Borel-Pad\'e re-summation of the $\beta$-function in the three Wigner-Dyson symmetry classes. Using this approximate $\beta$-function we discuss the dimensional dependence of the critical exponent and compare with numerical estimates. We also estimate the lower critical dimension of the symplectic symmetry class.}

\begin{document}
\maketitle
\newpage
\section{Introduction}

The Anderson transition is a disorder driven zero temperature quantum phase transition between a phase in which the electron motion is diffusive and a phase in which diffusion is suppressed, a phenomena known as Anderson localisation.\cite{Anderson58, Lee85, Evers08}
Similar to thermal phase transitions\cite{Nishimori11}, power law dependence of various physical quantities, referred to as critical phenomena, are observed in the vicinity of the transition.
The values of the critical exponents appearing in these power laws are expected to be universal, and to depend only on fundamental properties of the system, such as its dimensionality $d$ and the symmetries of its Hamiltonian.
For the Anderson transition the important symmetries are time reversal, spin rotation, chiral, and particle-hole.
There are ten symmetry classes\cite{Evers08}.
In this paper, we focus on the three Wigner-Dyson symmetry classes where chiral and particle-hole symmetries are absent.

For the Anderson transition there are two independent exponents $\nu$ and $z$.
The critical exponent $\nu$ describes the power law divergence of the correlation length at the transition, i.e.
\begin{equation}
	\xi = \left| x - x _c\right|^{-\nu} ,
\end{equation}
(where $x$ is the parameter that is varied to drive the transition, such as Fermi energy, disorder etc.)
The dynamical exponent $z$ appears in the power law describing the anomalous diffusion of a wavepacket at the transition, i.e. the second moment of an initially localized wavepacket grows with time $T$ as $T^{2/z}$.
For the Anderson transition it has been confirmed both numerically\cite{Ohtsuki97} and experimentally\cite{Chabe08} that $z=d$.
However, apart from fractals in the orthogonal symmetry class with dimensions very close to 2, the theoretical predictions for the critical exponent $\nu$
obtained using an $\epsilon$-expansion of the non-linear $\sigma$ model (NL$\sigma$M)\cite{Wegner79,Schafer80,Jungling80,Efetov80},
which is the field theory for the Anderson transition,
are in poor agreement with numerical results. This remains true even if the relevant series are re-summed using the Borel-Pad\'{e} method.

The $\epsilon$-expansion is a method that is expected to be reliable near two dimensions, i.e. as $\epsilon  \rightarrow 0$ where
\begin{equation}
  \epsilon = d-2 \;.
\end{equation}
In a previous paper\cite{Ueoka14}, we found that for the orthogonal symmetry class the agreement between the $\epsilon$-expansion and numerical simulation results at finite epsilon is considerably improved if the
$\epsilon$-expansion is supplemented by incorporating additional information about the limiting behaviour of the critical exponent in high dimensions, namely
\begin{equation}\label{eq:exponent_asymptotic}
  \nu \left( \epsilon \right) \rightarrow \frac{1}{2} \,\,\, (\epsilon \rightarrow   \infty) \;.
\end{equation}
In Ref. \citen{Ueoka14} we showed how to modify the Borel-Pad\'{e} re-summation of the series for the critical exponent to incorporate this behaviour.
One drawback of the method described in Ref. \citen{Ueoka14} is that we obtain information only
about the critical exponent and not about other quantities that are of interest, such as the critical
conductance, $\beta$-function etc.
Another is that it is not applicable to the symplectic universality class because no suitable series for the critical exponent is available.
Here, we show how to overcome these drawbacks by working directly with the series for the $\beta$-function obtained in the $\epsilon$-expansion.
The essence of our approach is again to supplement the $\epsilon$-expansion by incorporating additional information, i.e. Eq. (\ref{eq:exponent_asymptotic}).
This is done by re-summing the series for the $\beta$-function in such a way that its slope at its zero approaches two for $d\rightarrow\infty$.
Since the slope at the zero is the reciprocal of the critical exponent we recover Eq. (\ref{eq:exponent_asymptotic}).
We report the results of this approach for all the Wigner-Dyson classes.
This approach also has the advantage of concentrating attention on the dimensionality dependence of the $\beta$-function.
This suggests a simple way to estimate the lower critical dimension of the symplectic universality class, which is an imprtant open problem in the theory of Anderson localisation.

The paper is organized as follows.
In Sect. \ref{sec:beta} we recall the results of the $\epsilon$-expansion for the $\beta$-function of the Anderson localisation problem.
In Sect. \ref{sec:beta_resum}, we describe a re-summation for the $\beta$-function and its application
to the orthogonal, symplectic and unitary Wigner-Dyson symmetry classes.
In Sect. \ref{sec:discussion} we conclude.

\section{$\beta$-function}\label{sec:beta}

In the effective field theory for the Anderson transition, a central role is played by the $\beta$-function
\begin{equation}
  	\beta(t) = - \frac{\mathrm{d} t}{\mathrm{d} \ln L} \;.
\end{equation}
Here, $t(>0)$ is proportional to the inverse of a suitable average $g$ of the zero-temperature conductances of an ensemble of $d$ dimensional hypercubes of side $L$
\begin{equation} \label{eq:def_t}
	t = \frac{1}{\pi g} \;.
\end{equation}
Here, $g$ is in units of $e^2/h$ and includes the sum over spin degrees of freedom.
The $\beta$-function describes the renormalisation of the conductance with system size.
A zero of the $\beta$-function at some critical $t_c$
\begin{equation}\label{eq:def_tc}
	\beta(t_c) = 0 \;,
\end{equation}
indicates the occurrence of the Anderson transition at the corresponding critical conductance $g_c$.
The critical exponent $\nu$ is related to the derivative of the $\beta$-function at its zero
\begin{equation}\label{eq:beta_t_deriv}
  \left.\frac{\mathrm{d} \beta}{\mathrm{d} t} \right|_{t_c} = -\frac{1}{\nu} \;.
\end{equation}
In the scaling theory of localisation\cite{Abrahams79}, a different $\beta$-function, defined as
\begin{equation}
  \beta(g) = \frac{\mathrm{d} \ln g}{\mathrm{d} \ln L} \;,
\end{equation}
is used.
The two $\beta$-functions are related to each other by
\begin{equation}\label{eq:relation_beta}
\beta(g) = \frac{\beta(t)}{t} \;.
\end{equation}
Using this $\beta$-function the equation for the critical exponent becomes
\begin{equation}\label{eq:beta_g_deriv}
	\left. \frac{\mathrm{d} \beta\left(g\right)}{\mathrm{d} \ln g} \right|_{g_c} = \frac{1}{\nu} \;.
\end{equation}

Perturbation expansions for the beta functions have been calculated up to five loop order using the $\epsilon$-expansion method.\cite{Wegner89, Hikami92}
For orthogonal symmetry the result is
\begin{equation}\label{eq:beta_o}
  \beta (t) = \epsilon t - 2t^2 -12\zeta(3)t^5 + \frac{27}{2}\zeta(4)t^6 + \mathcal{O}(t^7) \;,
\end{equation}
for symplectic symmetry
\begin{equation}\label{eq:beta_s}
  \beta (t) = \epsilon t + t^2 -\frac{3}{4}\zeta(3)t^5 -\frac{27}{64}\zeta(4)t^6  + \mathcal{O}(t^7) \;,
\end{equation}
and for unitary symmetry
\begin{equation}\label{eq:beta_u}
  \beta (t) = \epsilon t - 2t^3 - 6t^5 + \mathcal{O}(t^7) \;.
\end{equation}
From these expansions, for the orthogonal and unitary symmetry classes, the dimensional dependence of the critical exponent as a series in powers of $\epsilon$
has been calculated by solving Eq. (\ref{eq:def_tc}) and then evaluating Eq. (\ref{eq:beta_t_deriv}).

\section{Re-summation of the $\beta$ function} \label{sec:beta_resum}

The central idea of our approach is to perform a Borel-Pad\'e re-summation of the $\beta$-function in such a way that the slope of the $\beta$-function at the critical point approaches a positive constant $A$ in high dimensions
\begin{equation}\label{eq:betagc}
	\left. \frac{\mathrm{d} \beta\left(g\right)}{\mathrm{d} \ln g} \right|_{g_c} \rightarrow A \;\;\; (\epsilon \rightarrow \infty) \;.
\end{equation}
Noting that all  the $\beta$-functions found in the $\epsilon$-expansion obey
\begin{equation}\label{eq:beta_g_eps_dep}
  \beta(g, \epsilon) = \epsilon + \beta(g, \epsilon = 0) \;,
\end{equation}
and since
\begin{equation}\label{eq:gcasymp}
 g_c \rightarrow 0 \,\,\, (\epsilon \rightarrow \infty) \;,
\end{equation}
re-summing the $\beta$-function so that
\begin{equation}\label{eq:betag2}
	\beta(g) \sim A \ln g \;\;\; (g \rightarrow 0) \;,
\end{equation}
will give the desired behaviour.
If we set $A=2$, this will give the correct behaviour for the critical exponent $\nu$ as $\epsilon \rightarrow \infty$.

The reader may at this point object that the expected behaviour of the $\beta$-function
in the strongly localised regime is
\begin{equation}
	\beta(g) \sim \ln g \;\;\; (g \ll g_c) \;,
\end{equation}
i.e, it would seem more natural to set $A=1$ in Eq. (\ref{eq:betag2}).
However,
if we set $A=1$, we will inevitably obtain
\begin{equation}\label{eq:exponent_asymptotic_a=1}
  \nu \left( \epsilon \right) \rightarrow 1 \,\,\, (\epsilon \rightarrow \infty) \;.
\end{equation}
which is not correct. As long as we rely on the results of the $\epsilon$-expansion, we are thus forced to compromise.
We set $A=2$.
The approximation for the $\beta$-function we obtain by doing so has by construction the
correct behaviour in the metallic regime and, as we shall demonstrate below by direct comparison with a $\beta$-function obtained from a finite size scaling (FSS) analysis of numerical simulation data, is a very reasonable approximation in the critical regime.
The price that must be paid is that our re-summation does not correctly describe the $\beta$-function in the strongly localised regime.
Since we focus here on the Anderson transition, i.e. the critical regime, this
is acceptable.

We now turn to the mechanics of how the re-summation is effected.
The results of the $\epsilon$-expansion for the $\beta$-functions of the Wigner-Dyson classes have the following common form\cite{Hikami92},
\begin{equation}
	\beta(t) = \epsilon t - t f(t) \;.
\end{equation}
Substituting this form into Eq. (\ref{eq:beta_t_deriv}) we find
\begin{equation}
   \left.t \frac{\mathrm{d} f}{\mathrm{d} t} \right|_{t_c} = \frac{1}{\nu} \;.
\end{equation}
Since $t_c\rightarrow \infty$ as $\epsilon \rightarrow \infty$, the desired re-summation is one such that
\begin{equation}\label{eq:h_limit}
	t \frac{\mathrm{d} f}{\mathrm{d} t} \rightarrow A \;\;\; (t \rightarrow \infty) \;.
\end{equation}
To effect such a re-summation we differentiate the series for $f$, multiply by $t$ and express the result in terms of a new function $h$
\begin{equation}\label{eq:defh}
    t\frac{\mathrm{d} f(t)}{\mathrm{d} t} = A + h(t) \;.
\end{equation}
We then apply the standard Borel-Pad\'e method\cite{Bender99} to $h$.
The first step is to form the Borel sum
\begin{equation}
  \tilde{h}\left( x \right) = \sum_j \frac{1}{j!}h_j x^j \;,
\end{equation}
from the series
\begin{equation}
  h\left( x \right) = \sum_j h_j x^j \;,
\end{equation}
for $h$  by dividing the coefficient of the $j$th power in the series by $j!$ for each $j$.
We then express the function $h$ as the Borel transform of $ \tilde{h}$
\begin{equation}
    h(t) = \frac{1}{t} \int_{0}^{\infty}dx e^{-x/t} \tilde{h} \left( x \right) \;.
\end{equation}
The next step is to approximate $\tilde{h}$ by an appropriate Pad\'e approximation $r$
\begin{equation}
  \tilde{h}\left( x \right) \approx r \left( x \right) = \frac{p \left( x \right)}{q \left( x \right)} \;.
\end{equation}
Here, $p$ is a polynomial of order $m$ and $q$ is a polynomial of order $n>m$. We then have the following approximation $H\left(t\right)$ for $h\left(t\right)$
\begin{equation}\label{eq:h_integral}
    h\left(t\right) \approx H\left(t\right) = \frac{1}{t} \int_{0}^{\infty}dx e^{-x/t} r \left( x \right) \;.
\end{equation}
Since $n>m$, the approximation  $H$  satisfies
\begin{equation}
	H(t) \rightarrow 0 \;\;\; (t \rightarrow \infty) \;,
\end{equation}
as required by Eq. (\ref{eq:h_limit}). To perform the integration we decompose the Pad\'{e} approximation $r(x)$ into partial fractions
\begin{equation}
   r(x) = \sum_{j=1}^{n} \frac{a_j}{x-\lambda_j} \;.
\end{equation}
Since $r(x)$ is a real function, if complex $a_j$ and $\lambda_j$ occur in the summation,
their complex conjugates must also appear in the summation.
The result of the integration is
 \begin{equation}\label{eq:h_sum}
   H\left(t\right) = \frac{1}{t} \sum_{j=1}^{n} a_j B\left(\lambda_j/t\right) \;.
 \end{equation}
The function $B$ is expressed using the
exponential integrals $\mathrm{Ei}(x)$ and $\mathrm{E_1}(z)$ for non-zero real and complex arguments respectively\cite{Jeffrey04}, as
\begin{equation}\label{eq:def_B}
	B(s) = \left \{
		\begin{array}{cl}
		 - \exp(-s) \mathrm{Ei} ( s ) & (s \in \mathbb{R} \;\; s \neq 0)\\
		 \exp(-s)  \mathrm{E}_1 \left( -s \right) & (s \in \mathbb{C}\; \;  $arg$s\ne\pi) \;.
   		\end{array} \right.
\end{equation}
To obtain an approximation $F\left(t\right)$ for $f\left(t\right)$ a further integration is needed
\begin{equation}\label{eq:f_from_h}
    f\left(t\right) \approx  F\left( t \right) = \int_{0}^{t} \frac{A+H\left(t\right)}{t}dt \;.
\end{equation}
The result can be expressed in the form
\begin{equation}\label{eq:f_sum}
F\left(t\right) = \sum_{j=1}^{n} c_j B \left( \lambda_j / t \right) \;,
\end{equation}
where
\begin{equation}
  c_j = \frac{a_j}{\lambda_j} \;.
\end{equation}
The $\beta$-function is then approximated as
\begin{equation}\label{eq:beta_resum}
	\beta(t) \approx \epsilon t - tF\left(t\right) \;.
\end{equation}
Some further details of the calculations are described in Appendix\ref{sec:details}.

To estimate $t_c$, we solve Eq. (\ref{eq:def_tc}) numerically. We then evaluate Eq. (\ref{eq:beta_t_deriv}) to estimate the critical exponent $\nu$.

\subsection{Application to the orthogonal symmetry class}
\begin{figure}[tb]
\begin{center}
\includegraphics[scale=0.6]{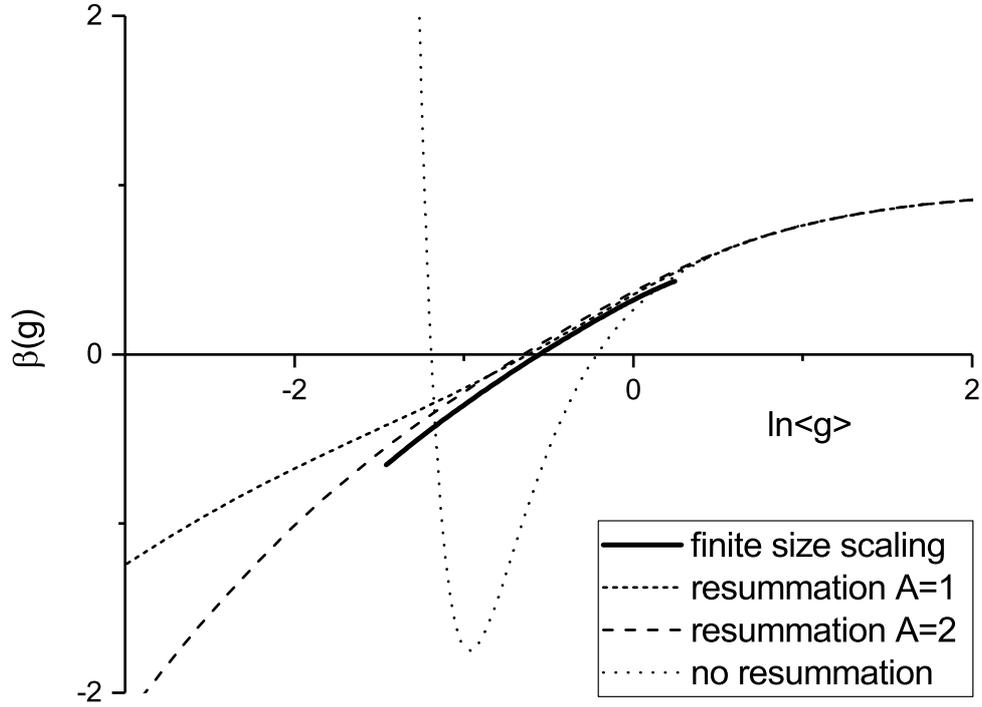}
\caption{Comparison of various approximations for the $\beta$-function for the $d=3$ orthogonal universality class with the $\beta$-function determined from finite size scaling (FSS) of numerical data obtained in simulations of Anderson's model of localisation.
Plotting of the FSS result is restricted to the conductance range for which numerical data is available.}
\label{plot:beta_oa}
\end{center}
\end{figure}

For the orthogonal symmetry class we have
\begin{equation}
  f(t) = 2t + 12\zeta(3)t^4 - \frac{27}{2} \zeta(4)t^5 + \mathcal{O}(t^6) \;,
\end{equation}
\begin{equation}
  h(t) = -2 + 2t + 48\zeta(3)t^4 - \frac{135}{2} \zeta(4)t^5 + \mathcal{O}(t^6) \;,
\end{equation}
\begin{equation}
  \tilde{h}(x) = -2+2x+2\zeta(3)x^4-\frac{9}{16}\zeta(4)x^5 +\mathcal{O}(x^6) \;.
\end{equation}
We use the $[2/3]$ Pad\'e approximation
\begin{equation}
  r(x) = \frac{-2+\left(2-\frac{9\zeta(4)}{16\zeta(3)}\right)x+\left(2\zeta(3)+\frac{9\zeta(4)}
  {16\zeta(3)}\right)x^2}{1+\frac{9\zeta(4)}{32\zeta(3)}x-\zeta(3)x^2-\zeta(3)x^3} \;.
\end{equation}
The number of independent coefficients in this Pad\'e approximation is the same as the number of known coefficients in the series and this approximation also satisfies the condition that $r$ goes to zero at infinity.
There are three terms in the partial fraction expansion of the Pad\'e approximation,
\begin{align}
	\lambda_1 &\simeq -0.8759+0.5824i \;, \nonumber \\
	\lambda_2 &= \lambda_1^{*}\;, \nonumber \\
	\lambda_3 &\simeq 0.7519\;, \nonumber \\
	c_1 &\simeq 1.1422 + 0.1773i\;, \nonumber \\
	c_2 &= c_1^{*}\;, \nonumber \\
	c_3 &\simeq -0.2844 \;.
\end{align}
The $\beta$-function is then approximated by Eq. (\ref{eq:beta_resum}).
We plot the corresponding approximation for the $\beta$-function of the scaling theory of localisation
for the orthogonal symmetry class for $d=3$ in Fig. \ref{plot:beta_oa}.
In the figure, the dotted line corresponds to the $\beta$ function obtained from the perturbation series
Eq. (\ref{eq:beta_o}) without re-summation.
The long and short dashed lines correspond, respectively, to the Borel-Pad\'e re-summation with $A=1$ and $A=2$.
The solid line was obtained from a finite size scaling analysis of data for the two-terminal Landauer conductance obtained in simulations of Anderson's model of localisation. (The details of the simulations are described in Appendix\ref{sec:fss}.)
In the metallic region, all the approximations agree as required.
In the critical region, the series without re-summation is a poor approximation, while
both the Borel-Pad\'e re-summations give reasonable approximations, with $A=2$ slightly closer to the
FSS result.
In the localised regime, we expect the $\beta$ function to approach a straight line with slope unity but it is evident that this regime will be reached only at much smaller conductance.

\begin{figure}[tb]
\begin{center}
\includegraphics[scale=0.6]{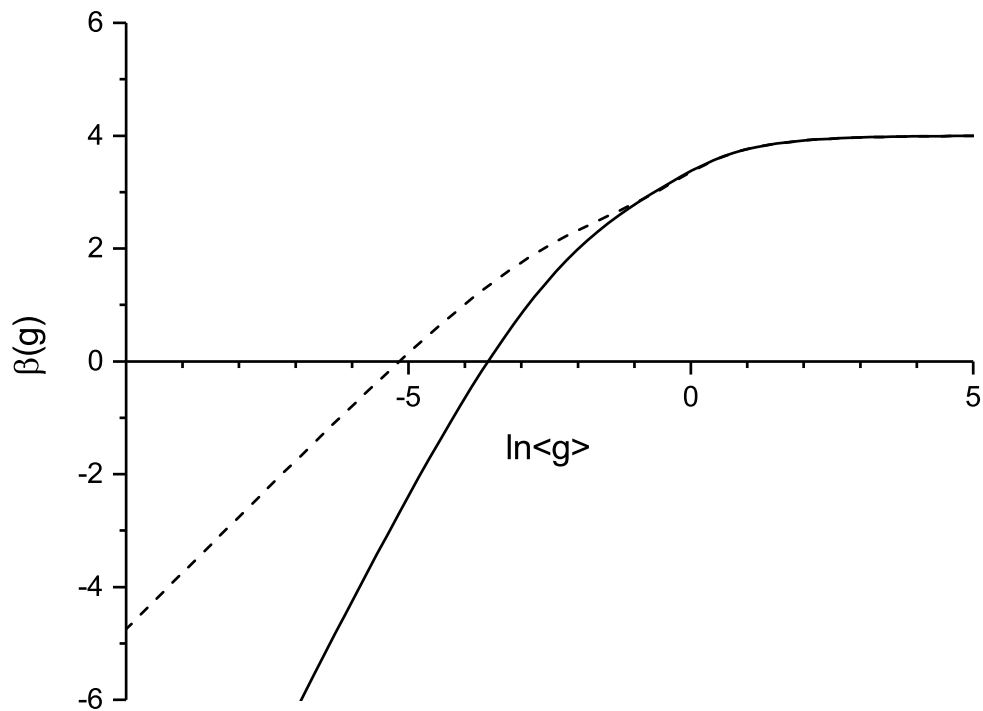}
\caption{Comparison of the approximations with $A=1$ (dashed line) and $A=2$ (solid line) for the $\beta$-function for the $d=6$ orthogonal universality class.}
\label{plot:beta_ob}
\end{center}
\end{figure}

\begin{table}[ht]
	\begin{center}
	\begin{tabular}{|c||c|c||c||c|} \hline
            &                       &       &  Ref. \citen{Ueoka14} &  numerical estimate    \\
      	$d$ & $g_c$                 & $\nu$ & $\nu$ & $\nu$                                  \\ \hline
      	$3$ & $5.23\times 10^{-1}$  & 1.64  & 1.46  & $1.571 \pm .004$ Ref. \citen{Slevin14} \\ \hline
      	$4$ & $1.37\times 10^{-1}$  & 1.06  & 1.06  & $1.156 \pm .014$ Ref. \citen{Ueoka14}  \\ \hline
      	$5$ & $5.60\times 10^{-2}$  & 0.775 & 0.891 & $0.969 \pm .015$ Ref. \citen{Ueoka14}  \\ \hline
      	$6$ & $2.76\times 10^{-2}$  & 0.655 & 0.798 & $0.78  \pm .06$ Ref. \citen{Garcia07}  \\ \hline
	\end{tabular}
	\end{center}
	\caption{
	Approximate fixed points $g_c$ and critical exponents for the orthogonal symmetry class for $d=3,4,5$ and $6$ obtained using a re-summation of the $\beta$-function with $A=2$.
 The approximations for the critical exponents obtained in Ref. \citen{Ueoka14} as well as available numerical estimates are also listed.
} \label{table:o_integer_d}
\end{table}

\begin{table}[ht]
	\begin{center}
	\begin{tabular}{|c|c|c|} \hline
      	$d$ & $g_c$                & $\nu$  \\ \hline
      	$3$ & $5.39\times 10^{-1}$ & 1.81  \\ \hline
      	$4$ & $7.38\times 10^{-2}$ & 1.71  \\ \hline
      	$5$ & $1.80\times 10^{-2}$ & 1.22  \\ \hline
      	$6$ & $5.75\times 10^{-3}$ & 1.09  \\ \hline
	\end{tabular}
	\end{center}
	\caption{Approximate fixed points $g_c$ and critical exponents for the orthogonal symmetry class for $d=3,4,5$ and $6$
obtained using a re-summation of the $\beta$-function with $A=1$.
} \label{table:o_integer_d_a=1}
\end{table}

In Table \ref{table:o_integer_d} we list the approximations for the critical exponent obtained with $A=2$ for $d = 3,4,5$ and $6$ and compare these with numerical estimates obtained using FSS.
For  $d=3$ and $4$ the agreement is reasonable, but for $d=5$ and $6$ the agreement is slightly worse than the method of Ref. \citen{Ueoka14}.

In Table \ref{table:o_integer_d_a=1}, we list the approximations for the critical exponent obtained with
$A=1$.
Comparing with Table \ref{table:o_integer_d}, it's clear that re-summation with $A=1$ gives much poorer approximations for the critical exponent.
In Fig. \ref{plot:beta_ob}, we compare the $\beta$-functions obtained using the Borel-Pad\'e re-summations in $d=6$ with $A=1$ and $A=2$.
In the strongly localised regime both approximations approach straight lines with slope $A$, i.e. slopes 1 and 2 for $A=1$ and $A=2$, respectively.
Given that the $\beta$-functions satisfy Eq. (\ref{eq:beta_g_eps_dep}), i.e. that varying the dimension simply translates the curve for the $\beta$-function in the direction of the ordinate,
the figure makes clear that re-summation with $A=1$ cannot approximate the behaviour of the $\beta$-function near  the critical point in high dimensions.

\subsection{Application to the symplectic symmetry class}

\begin{figure}[tb]
\begin{center}
\includegraphics[scale=0.6]{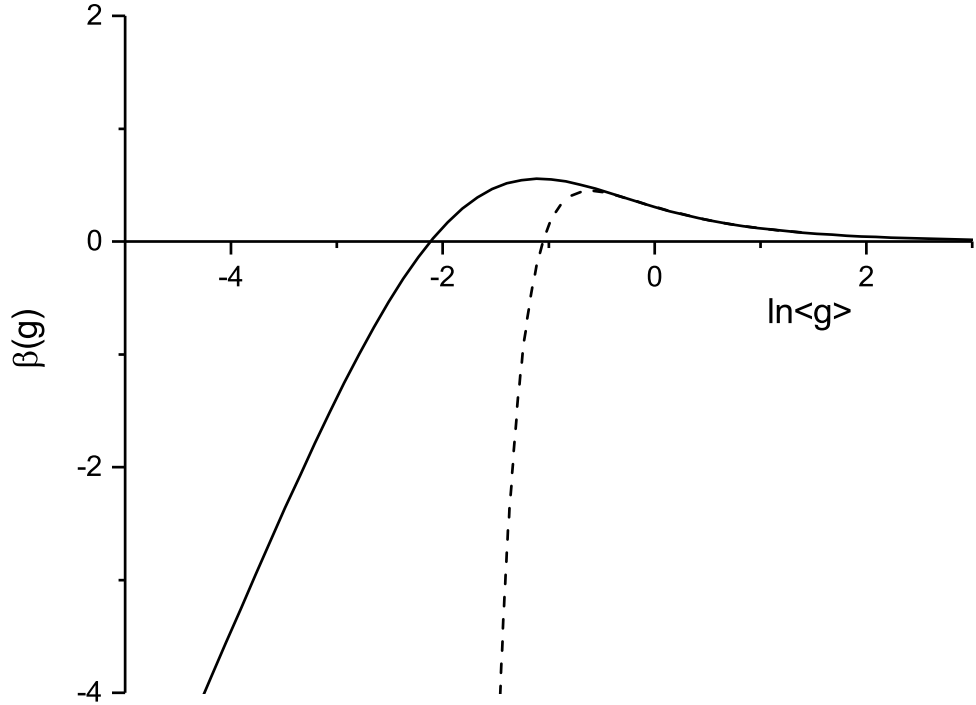}
\caption{The approximation for the $\beta$-function for the $d=2$ symplectic universality class. The dashed line is the approximation obtained from the $\epsilon$-expansion without re-summation.}
\label{plot:beta_sa}
\end{center}
\end{figure}

The $\beta$ functions of the orthogonal and symplectic symmetry classes are related by \cite{Hikami92}
\begin{align}
	\beta_S(t)=-2\beta_O(-\frac{t}{2}).
\end{align}
Using this relationship and the re-summation for the orthogonal symmetry class the approximation for the $\beta$-function for the symplectic symmetry class for  dimension $d=2$ is plotted in Fig. \ref{plot:beta_sa}.
The estimates of the critical exponent for various dimensions are listed in Table. \ref{table:s_integer_d}.

Unfortunately the results for the critical exponent are disappointing.
For $d=2$ we find $\nu\approx 0.87$.
Moroz\cite{Moroz96} proposed the $\mu$-Pad\'e method and applied it to the symplectic symmetry class. He reports, $\nu \approx 0.98$ for $d=2$.
Neither value compares well with the numerical estimate $\nu\approx 2.73$.
For $d=3$, the known lower-bound on the value of the critical exponent\cite{Chayes86, Kramer93}
\begin{equation}
	\nu \ge \frac{2}{d} \label{eq:lowerbound} \;,
\end{equation}
is violated
and the agreement with the numerical estimate\cite{Asada05} is poor.
For higher dimensions the approach to the asymptotic value is non-monotonic.
A possible cause of these difficulties may be that, compared with the orthogonal and unitary symmetry classes,
more terms in the series for the $\beta$-function  are required to accurately describe the non-monotonic behaviour of the $\beta$-function that is special to the symplectic symmetry class.

\begin{table}[ht]
	\begin{center}
	\begin{tabular}{|c||c|c||c|} \hline
      	$d$ & $g_c$ & $\nu$ & Ref.\cite{Asada05,Asada02} \\ \hline
      	$2$ & $1.2  \times 10^{-1}$ &$0.874$ & $2.73\pm.02$ \\ \hline
      	$3$ & $6.13 \times 10^{-2}$ &$0.565$ & $1.375\pm.008$ \\ \hline
      	$4$ & $3.64 \times 10^{-2}$ &$0.492$ & \\ \hline
      	$5$ & $2.26 \times 10^{-2}$ &$0.466$ & \\ \hline
      	$6$ & $1.42 \times 10^{-2}$ &$0.459$ & \\ \hline
	\end{tabular}
	\end{center}
	\caption{
	Approximations for the critical exponents and fixed points $g_c$ for the symplectic symmetry class from $d=2$--$6$.
For $d\ge 20$ we find $\nu=1/2$ to the accuracy shown.
Available numerical estimates are also listed.} \label{table:s_integer_d}
\end{table}

\begin{figure}[tb]
\begin{center}
\includegraphics[scale=0.6]{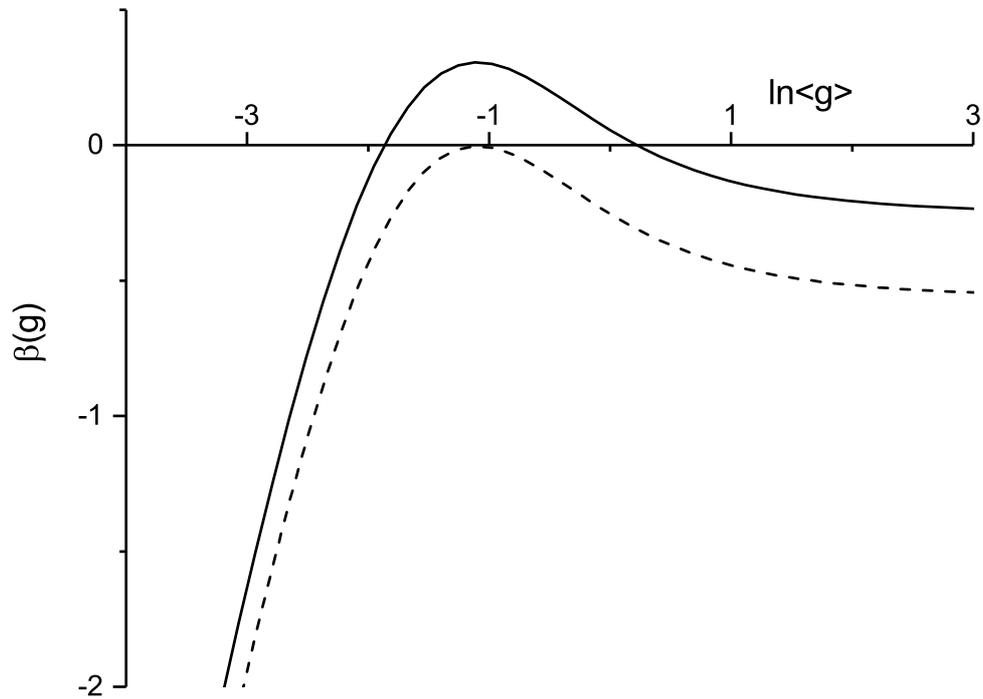}
\caption{The approximation for the $\beta$ function for the symplectic symmetry class for $d=1.75$ (solid line) and $d=1.44$ (dashed line).}
\label{plot:beta_s_dl}
\end{center}
\end{figure}

As shown in Fig. \ref{plot:beta_s_dl}, for the symplectic symmetry class for dimensions below two, and above the lower critical dimension $d_l$, two fixed points appear:
a critical fixed point and a stable fixed point.
Numerical simulations on fractals have been reported that support the existence of both these
fixed points.\cite{Asada06,Sticlet16}
At the lower critical dimension these fixed points annihilate each other and the value of the $\beta$-function at its maximum is zero
\begin{equation}
  \max \beta(g) = 0 \;.
\end{equation}
Making use of Eq. (\ref{eq:beta_g_eps_dep}) we then obtain the following estimate for the lower critical dimension of the symplectic symmetry class
\begin{equation}
  d_l \approx 2 - \max  \beta_{S} \left( g , \epsilon = 0 \right) \;.
\end{equation}
Treating the second term numerically, we find
 \begin{equation}\label{eq:d_l}
   d_l \simeq 1.44 \;.
 \end{equation}
If we perform the same calculation without using the Borel-Pade method to re-sum the series we obtain
 \begin{equation}
   d_l \simeq 1.55 \;.
 \end{equation}
These predictions remain to be confirmed in future numerical work.

\subsection{Application to the unitary symmetry class}
\begin{figure}[tb]
\begin{center}
\includegraphics[scale=0.6]{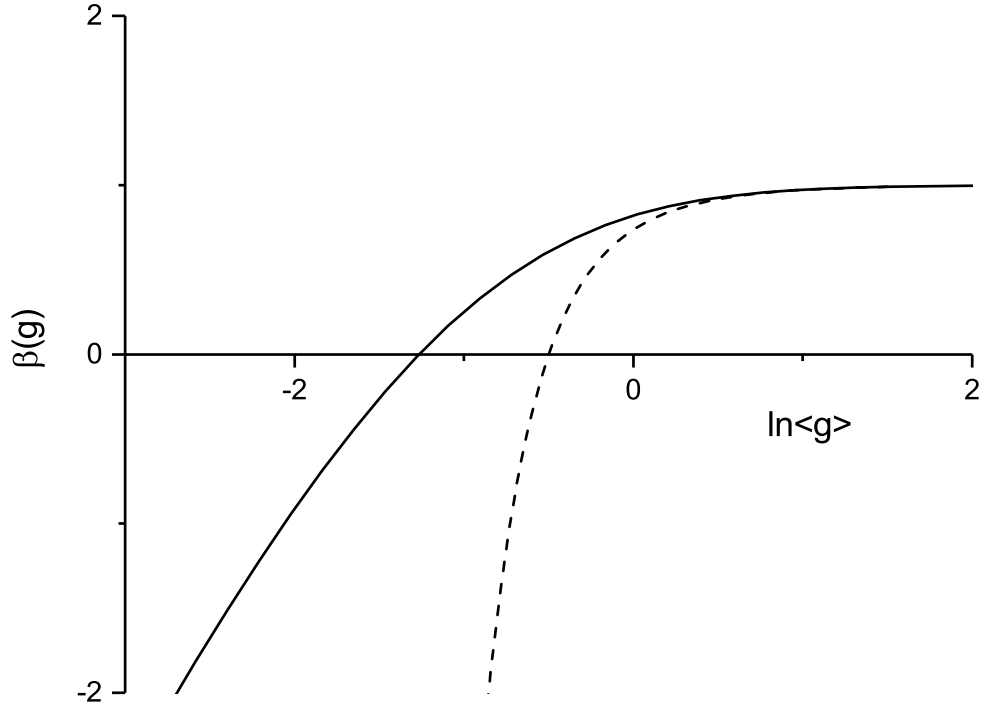}
\caption{The approximation for the $\beta$-function in unitary $d=3$ universality class.
The dashed line is the approximation obtained from the $\epsilon$-expansion without re-summation.}
\label{plot:beta_u}
\end{center}
\end{figure}

For the unitary symmetry class we have
\begin{equation}
  f(t) = 2t^2 + 6t^4 + \mathcal{O}(t^6),
\end{equation}
\begin{equation}
  h(t) = -2 +4t^2 + 24t^4 + \mathcal{O}(t^6),
\end{equation}
\begin{equation}
  \tilde{h}(x) = -2 + 2x^2 + x^4 +\mathcal{O}(x^6).
\end{equation}
We use the $[0/4]$ Pad\'e approximation
\begin{equation}
  r(x) = -\frac{4}{3x^4+2x^2+2} \;.
\end{equation}
This Pad\'e approximation has three independent coefficients which matches the number of known coefficients in the original series. (In counting coefficients we consider even power of $x$ only.)
There are four terms in the partial fraction expansion of the Pad\'e approximation
\begin{align}
	\lambda_1 &\simeq -0.4915+0.7582i\;, \nonumber \\
	\lambda_2 &= \lambda_1^{*}\;, \nonumber \\
	\lambda_3 &= -\lambda_1\;, \nonumber \\
	\lambda_4 &= \lambda_3^{*}\;, \nonumber \\
	c_1 &\simeq 0.5 + 0.2236i\;, \nonumber \\
	c_2 &= c_1^{*}\;, \nonumber \\
	c_3 &= c_1\;, \nonumber \\
	c_4 &= c_3^{*} \;.
\end{align}
The approximation for the $\beta$-function is plotted for dimension $d=3$ in Fig. \ref{plot:beta_u}.
Approximations for the critical exponent for $d=3,4,5$ and $6$ are listed in Table. \ref{table:u_integer_d}.
We also list the estimates of the critical exponent obtained from the standard Borel-Pad\'e re-summation and the Borel-Pad\'e re-summation following the method of Ref. \citen{Ueoka14}. (The details of the relevant calculations are given in Appendix\ref{sec:previous_method}.)
The values obtained using the method of Ref. \citen{Ueoka14}, Eq. (\ref{eq:prevunitary}), and the re-summation of the $\beta$-function
proposed here, both obey the lower bound Eq. (\ref{eq:lowerbound}) for the critical exponent
which is not the case for the standard Borel-Pad\'e re-summation Eq. (\ref{eq:stdunitary}).
Re-summation of the $\beta$-function leads to a further minor improvement in the agreement with numerical estimates compared with the method of Ref. \citen{Ueoka14} but the disagreement with numerical estimate is still significant.

\begin{table}[ht]
	\begin{center}
	\begin{tabular}{|c||c|c||c|c||c|} \hline
      	$d$ & Eq.(\ref{eq:stdunitary}) & Eq.(\ref{eq:prevunitary}) & $g_c$ & $\nu$ & Ref. \cite{Ujfalusi15a,Ujfalusi15b, Slevin16} \\ \hline
      	$3$ & 0.26 & 0.71 & $2.82\times 10^{-1}$ & 0.969 & $1.437[.426,.448]$ \\ \hline
      	$4$ & 0.1  & 0.58 & $1.27\times 10^{-1}$ & 0.687 & $1.1[.09,.12]$ \\ \hline
      	$5$ & 0.05 & 0.54 & $6.76\times 10^{-2}$ & 0.595 & \\ \hline
      	$6$ & 0.03 & 0.53 & $3.82\times 10^{-2}$ & 0.552 & \\ \hline
	\end{tabular}
	\end{center}
	\caption{
	The approximations for the critical exponent $\nu$ obtained with the standard Borel-Pad\'e re-summation Eq. (\ref{eq:stdunitary}), the Borel-Pad\'e re-summation following the method of Ref. \citen{Ueoka14}, i.e. Eq. (\ref{eq:prevunitary}), and the Borel-Pad\'e re-summation of the $\beta$-function, described in Sect. \ref{sec:beta_resum}, for the unitary symmetry class for $d=3,4,5$ and $6$. For the latter, the estimates of the fixed points $g_c$ are also given. The available numerical estimates are also listed.
	} \label{table:u_integer_d}
\end{table}

\section{Discussion}\label{sec:discussion}

We have suggested a Borel-Pad\'e re-summation of the $\beta$-function for the Anderson localisation
problem in the Wigner-Dyson symmetry classes.
This work extends and complements the approach in Ref. \citen{Ueoka14}.
By focusing on the $\beta$-function we were able to make predictions for quantities such as, for example, the critical conductance that were not possible when working with the series for the critical exponent.
We were also able to make a prediction for the lower critical dimension of the symplectic symmetry class.
It should be possible to check these predictions in future numerical simulations.

Ideally, we would like to be able to re-sum the series for the $\beta$-function so that it has the correct form and dimensionality dependence not only in the metallic regime and at the critical point but also in the localised regime.
While this problem remains to be solved, we think the general idea of supplementing the $\epsilon$-expansion, which is valid near two dimensions, with additional results for higher dimensions and interpolating between them is promising.

\section*{Acknowledgements}
We would like to thank Kazuhiko Kuroki, Kensuke Kobayashi, Hikaru Kawamura and Kenichi Asano for their comments and suggestions.
This work was supported by JSPS KAKENHI Grants Numbers JP15H03700 and 26400393.

\bibliography{references}

\begin{thebibliography}{10}

\bibitem{Anderson58}
P.~W. Anderson: Phys. Rev. {\bfseries 109} (1958) 1492.

\bibitem{Lee85}
P.~A. Lee and T.~V. Ramakrishnan: Rev. Mod. Phys. {\bfseries 57} (1985) 287.

\bibitem{Evers08}
F.~Evers and A.~D. Mirlin: Rev. Mod. Phys. {\bfseries 80} (2008) 1355.

\bibitem{Nishimori11}
H.~Nishimori and G.~Ortiz: {\em Elements of phase transitions and critical
  phenomena} (Oxford graduate texts. Oxford University Press, New York, 2011),
  Oxford graduate texts.

\bibitem{Ohtsuki97}
T.~Ohtsuki and T.~Kawarabayashi: Journal of the Physical Society of Japan
  {\bfseries 66} (1997) 314.

\bibitem{Chabe08}
J.~Chab\'{e}, G.~Lemari\'{e}, B.~Gr\'{e}maud, D.~Delande, P.~Szriftgiser, and
  J.~C. Garreau: Physical Review Letters {\bfseries 101} (2008) 255702.

\bibitem{Wegner79}
F.~Wegner: Z. Phys. B {\bfseries 35} (1979) 207.

\bibitem{Schafer80}
L.~Sch{\"{a}}fer and F.~Wegner: Z. Phys. B {\bfseries 38} (1980) 113.

\bibitem{Jungling80}
K.~J\"ungling and R.~Oppermann: Z. Phys. B {\bfseries 38} (1980) 93.

\bibitem{Efetov80}
K.~B. Efetov, A.~I. Larkin, and Kheml'nitskii: Sov. Phys. JETP {\bfseries 52}
  (1980) 568.

\bibitem{Ueoka14}
Y.~Ueoka and K.~Slevin: Journal of the Physical Society of Japan {\bfseries 83}
  (2014) 084711.

\bibitem{Abrahams79}
E.~Abrahams, P.~W. Anderson, D.~C. Licciardello, and T.~V. Ramakrishnan: Phys.
  Rev. Lett. {\bfseries 42} (1979) 673.

\bibitem{Wegner89}
F.~Wegner: Nuclear Physics B {\bfseries 316} (1989) 663.

\bibitem{Hikami92}
S.~Hikami: Progress of Theoretical Physics Supplement {\bfseries 107} (1992)
  213.

\bibitem{Bender99}
C.~M. Bender and S.~A. Orszag: {\em Advanced mathematical methods for
  scientists and engineers} (Springer, New York, 1999).

\bibitem{Jeffrey04}
A.~Jeffrey: {\em Handbook of mathematical formulas and integrals} (Elsevier
  Academic Press, Amsterdam ; Boston, 2004) 3rd ed.

\bibitem{Slevin14}
K.~Slevin and T.~Ohtsuki: New Journal of Physics {\bfseries 16} (2014) 015012.

\bibitem{Garcia07}
A.~M. Garc{\'{\i}}a-Garc{\'{\i}}a and E.~Cuevas: Phys. Rev. B {\bfseries 75}
  (2007) 174203.

\bibitem{Moroz96}
A.~Moroz: Journal of Physics A: Mathematical and General {\bfseries 29} (1996)
  289.

\bibitem{Chayes86}
J.~T. Chayes, L.~Chayes, D.~S. Fisher, and T.~Spencer: Phys. Rev. Lett.
  {\bfseries 57} (1986) 2999.

\bibitem{Kramer93}
B.~Kramer: Phys. Rev. B {\bfseries 47} (1993) 9888.

\bibitem{Asada05}
Y.~Asada, K.~Slevin, and T.~Ohtsuki: J. Phys. Soc. Japan Suppl. {\bfseries 74}
  (2005) 238.

\bibitem{Asada02}
Y.~Asada, K.~Slevin, and T.~Ohtsuki: Phys. Rev. Lett. {\bfseries 89} (2002)
  256601.

\bibitem{Asada06}
Y.~Asada, K.~Slevin, and T.~Ohtsuki: Phys. Rev. B {\bfseries 73} (2006) 041102.

\bibitem{Sticlet16}
D.~Sticlet and A.~Akhmerov: Physical Review B {\bfseries 94} (2016) 161115.

\bibitem{Ujfalusi15a}
L.~Ujfalusi and I.~Varga: Phys. Rev. B {\bfseries 91} (2015) 184206.

\bibitem{Ujfalusi15b}
L.~Ujfalusi, M.~Giordano, F.~Pittler, T.~G. Kov\'acs, and I.~Varga: Phys. Rev.
  D {\bfseries 92} (2015) 094513.

\bibitem{Slevin16}
K.~Slevin and T.~Ohtsuki: Journal of the Physical Society of Japan {\bfseries
  85} (2016) 104712.

\bibitem{Slevin01}
K.~Slevin, P.~Markos, and T.~Ohtsuki: Physical Review Letters {\bfseries 86}
  (2001) 3594.

\bibitem{Slevin03}
K.~Slevin, P.~Markos, and T.~Ohtsuki: Physical Review B {\bfseries 67} (2003)
  155106.

\bibitem{Slevin00}
K.~Slevin, T.~Ohtsuki, and T.~Kawarabayashi: Physical Review Letters {\bfseries
  84} (2000) 3915.

\end{thebibliography}

\appendix

\section{Calculation of Eqs. (\ref{eq:h_integral}) and (\ref{eq:f_from_h})}\label{sec:details}

We evaluate Eq. (\ref{eq:h_integral}) by replacing the Pad\'{e} approximation with it's partial fraction expansion and evaluating term by term.
For $\lambda \in \mathbb{R} \setminus \{0\}$
\begin{equation}
  \int_{0}^{\infty} dx \frac{e^{-x/t}}{x-\lambda} =
  e^{-\lambda/t}  \int_{-\lambda/t}^{\infty} dy \frac{e^{-y}}{y} =
  - e^{-\lambda/t} \mathrm{Ei}(\lambda/t) = B \left( \lambda/t \right)\;.
\end{equation}
The calculation for $\lambda \in \mathbb{C} \setminus \mathbb{R}$ is similar.

We evaluate Eq. (\ref{eq:f_from_h}) as follows
\begin{equation}
\int_0^t {\frac{{H\left( t \right)}}{t}} dt = \sum\limits_{j=1}^{n} {{a_j}\int_0^t {\frac{1}{{{t^2}}}B\left( {\frac{{{\lambda _j}}}{t}} \right)dt} } \;.
\end{equation}
Evaluating term by term, for $\lambda \in \mathbb{R} \setminus \{0\}$
\begin{equation}
  \int_0^t {\frac{1}{{{t^2}}}B\left( {\frac{\lambda }{t}} \right)dt}
  =
   - \int_0^t {\frac{1}{{{t^2}}}{e^{ - {\lambda  \mathord{\left/
 {\vphantom {\lambda  t}} \right.
 \kern-\nulldelimiterspace} t}}}{\rm{Ei}}\left( {\frac{\lambda }{t}} \right)dt} \;.
\end{equation}
Integrating by parts we obtain
\begin{equation}
\left. { - \frac{1}{\lambda }{e^{ - {\lambda  \mathord{\left/
 {\vphantom {\lambda  t}} \right.
 \kern-\nulldelimiterspace} t}}}{\rm{Ei}}\left( {\frac{\lambda }{t}} \right)} \right|_0^t + \int_0^t {\frac{1}{\lambda }{e^{ - {\lambda  \mathord{\left/
 {\vphantom {\lambda  t}} \right.
 \kern-\nulldelimiterspace} t}}}\frac{d}{{dt}}{\rm{Ei}}\left( {\frac{\lambda }{t}} \right)dt} \;.
\end{equation}
Since
\begin{equation}
  \frac{d}{{dx}}{\rm{Ei}}\left( x \right) = \frac{{{e^x}}}{x} \;,
\end{equation}
we can simplify the second term
\begin{equation}
  \int_0^t {\frac{1}{{{t^2}}}B\left( {\frac{\lambda }{t}} \right)dt}
 =
  - \frac{1}{\lambda }{e^{ - {\lambda  \mathord{\left/
 {\vphantom {\lambda  t}} \right.
 \kern-\nulldelimiterspace} t}}}{\rm{Ei}}\left( {\frac{\lambda }{t}} \right) - \frac{1}{\lambda }\int_0^t {\frac{1}{t}dt} \;.
\end{equation}
Collecting terms
\begin{equation}
  \int_0^t {\frac{{H\left( t \right)}}{t}} dt
  =
  \sum\limits_j^{} {\frac{{{a_j}}}{{{\lambda _j}}}B\left( {\frac{{{\lambda _j}}}{t}} \right)}  - \sum\limits_{j = 1}^n {c_j\int_0^t {\frac{1}{t}dt} } \;.
\end{equation}
The calculation for $\lambda \in \mathbb{C} \setminus \mathbb{R}$ is similar.
Since
\begin{equation}
  \left. t\frac{df}{dt} \right|_{t=0} = 0 \;.
\end{equation}
we have
\begin{equation}
  h(0) = -A \;.
\end{equation}
From this we derive a sum rule
\begin{equation}
  A = -h(0) = -\tilde{h}(0) = -r(0)
  =
  \sum\limits_{j=1}^{n} { c_j } \;.
\end{equation}
We thus arrive at
\begin{equation}
  \int_0^t {\frac{{H\left( t \right)}}{t}} dt
  =
  \sum\limits_{j=1}^{n} {\frac{{{a_j}}}{{{\lambda _j}}}B\left( {\frac{{{\lambda _j}}}{t}} \right)}  - A\int_0^t {\frac{1}{t}dt} \;.
\end{equation}
This leads directly to Eq. (\ref{eq:f_sum}).

\section{Estimation of the $\beta$-function by finite size scaling}\label{sec:fss}

\begin{figure}[htb]
\begin{center}
\includegraphics[scale=0.6]{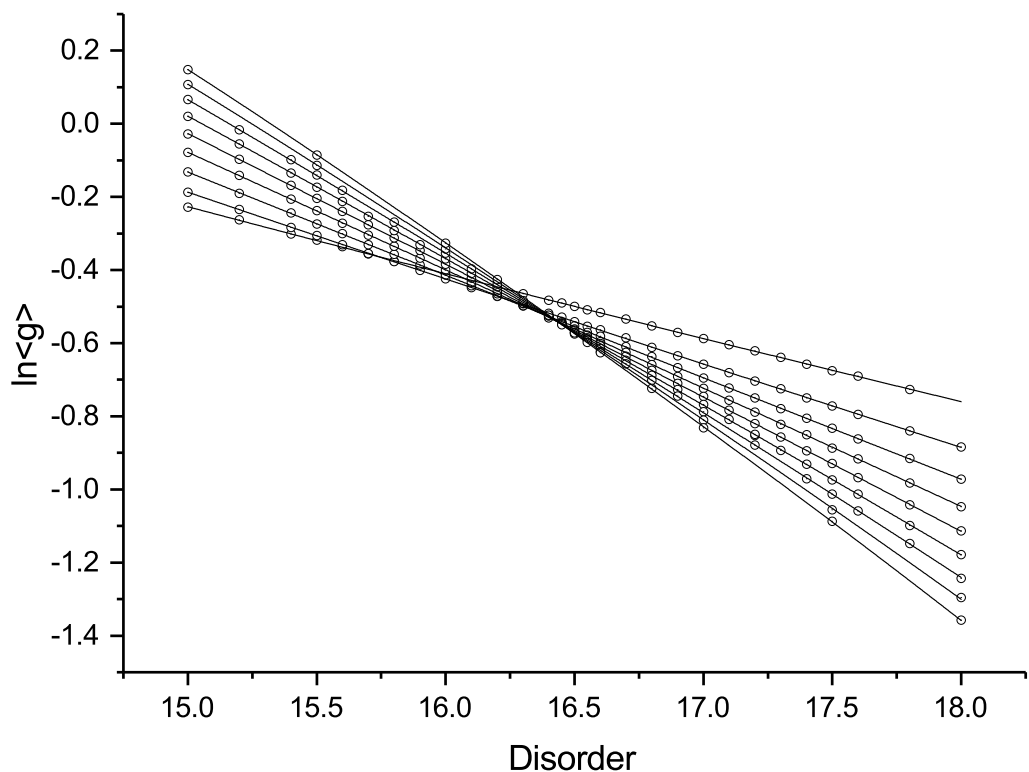}
\caption{Finite size scaling fit (solids lines) to conductance data (circles) for Anderson's model of localisation in three dimensions. The errors of the data points are smaller than the symbol size.}
\label{plot:fss_a}
\end{center}
\end{figure}

\begin{figure}[htb]
\begin{center}
\includegraphics[scale=0.6]{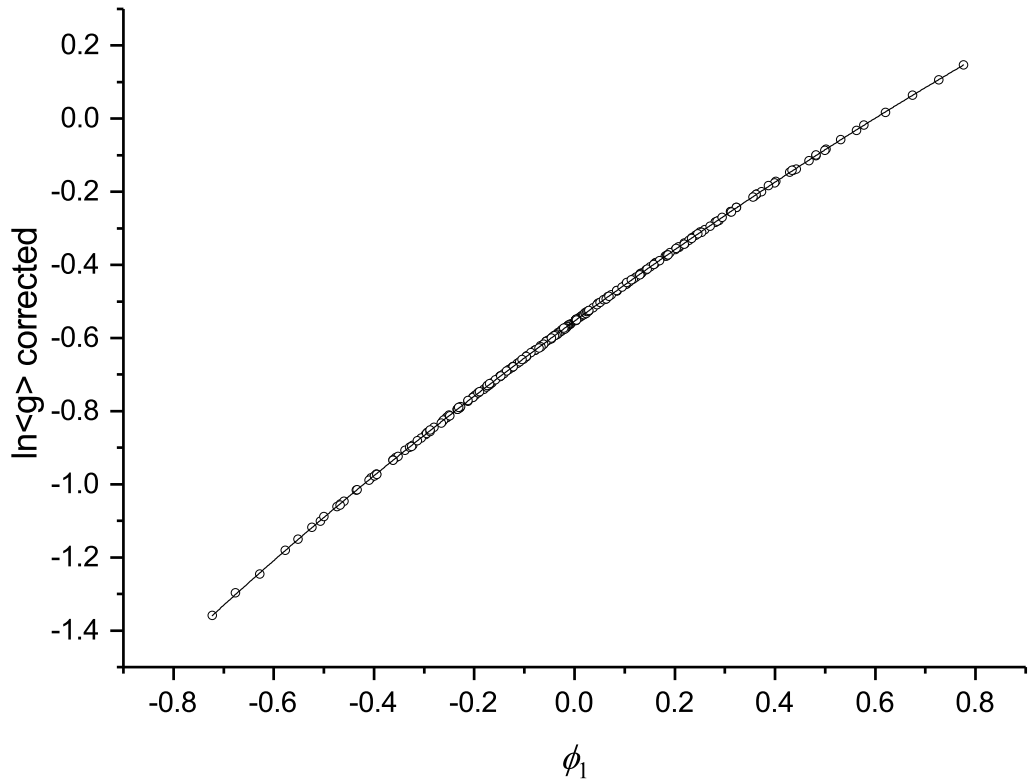}
\caption{Demonstration of single parameter scaling for the conductance data. As explained in the text, the correction due to the irrelevant scaling variable has been subtracted from the simulation data. The solid line is $F\left( \phi_1, 0 \right)$}.
\label{plot:fss_b}
\end{center}
\end{figure}

We performed a finite size scaling analysis of data obtained in previous
numerical simulations\cite{Slevin01,Slevin03} of Anderson's model of localisation.
In these simulations, ensembles of systems of size $L \times L \times L$ were simulated for various disorders $W$ and the zero temperature two terminal conductance $g$ calculated for a fixed Fermi energy $E_F=0.5$.
Fixed boundary conditions were imposed in the transverse directions.
(Note that the critical conductance distribution depends quite strongly on the choice of transverse boundary conditions.\cite{Slevin00})
As described in Refs. \citen{Slevin01} and \citen{Slevin03} an appropriate contact resistance has been subtracted from the resistance of each sample (see Eq. (5) of Ref. \citen{Slevin01}).
The data used in Refs. \citen{Slevin01} and \citen{Slevin03} has been supplemented by some additional data for some larger system sizes $L=18$ and $L=20$.
The precision of the data is typically a few tenths of a percent.
We then fitted the logarithm of the average of the conductance to
\begin{equation}
  \ln \left\langle g \right\rangle  = F\left( {{\phi _1},{\phi _2}} \right) \;,
\end{equation}
where
\begin{equation}
\begin{array}{*{20}{c}}
{{\phi _1} = {u_1}\left( w \right){L^\alpha }}&{{\phi _2} = {u_2}\left( w \right){L^{ y}}}&{w = W - {W_c}} \;.
\end{array}
\end{equation}
Here, $\alpha$ is the reciprocal of the critical exponent $\nu$
\begin{equation}
  \alpha = \frac{1}{\nu} \;,
\end{equation}
and $y<0$ is an irrelevant exponent.
The scaling function $F$ was expanded as a Taylor series in its arguments, and the functions $u_1$ and
$u_2$ were expanded in Taylor series in powers of $w$ subject to appropriate constraints.
The fitting procedure is described in detail in Ref. \citen{Slevin14}.
The number of data points was $222$, the number of parameters $14$ and the goodness of fit probability $0.5$.
The scaling function $F$ was expanded to 4th and 1st order in its 1st and 2nd arguments, respectively.
The function $u_1$ was expanded to 2nd order in $w$, and $u_2$ was truncated at a constant.
From the fit  we estimated $W_c= 16.457 \pm .005$, the critical value of $\ln\left<g\right> = -0.553 \pm .002$ and the critical exponent $\nu=1.587 \pm .006$.
All errors are standard errors and were calculated using Monte Carlo simulation with 250 samples.
The estimate of the critical exponent is consistent with previous estimates\cite{Slevin14} but there is small
$\approx 0.5 \%$ but statistically significant deviation from the critical disorder estimated using the transfer matrix method
which for the moment we cannot explain.
The data together with the finite size scaling fit are shown in Fig. \ref{plot:fss_a}.
After correcting the data
\begin{equation}
\ln \left< g \right> \mathrm{corrected} = \ln \left< g \right> - \Delta\left(W, L\right) \;,
\end{equation}
by subtracting the irrelevant correction
\begin{equation}
\Delta\left(W, L\right) = F\left( \phi_1, \phi_2 \right) - F\left( \phi_1, 0 \right) \;,
\end{equation}
the data should collapse onto a single curve. This is demonstrated in Fig. \ref{plot:fss_b}.

We obtain the $\beta$-function from the finite size scaling fit by plotting
\begin{equation}
\frac{{d\ln \left\langle g \right\rangle }}{{d\ln L}} = \frac{{dF\left( {{\phi _1},0} \right)}}{{d{\phi _1}}}\frac{{\partial {\phi _1}}}{{\partial L}}\frac{{dL}}{{d\ln L}} = \alpha {\phi _1}F'\left( {{\phi _1},0} \right) \;,
\end{equation}
versus
\begin{equation}
\ln \left\langle g \right\rangle  = F\left( {\phi_1,0} \right) \;.
\end{equation}

\section{Re-summation of the series for the exponent for the unitary symmetry class}\label{sec:previous_method}

Here, we describe the application of the method of Ref. \citen{Ueoka14} to the unitary symmtery class.
For the unitary symmetry class the $\epsilon$-expansion for the critical exponent is
\begin{equation}
  \nu = \frac{1}{2 \epsilon} - \frac{3}{4} + O(\epsilon) \;.
\end{equation}
Note that the unknown terms start at $O(\epsilon)$ as compared with $O(\epsilon^3)$ for the orthogonal case.
The standard Borel-Pad\'e re-summation is
\begin{equation}\label{eq:stdunitary}
   \nu  \approx \frac{1}{\epsilon^2} \int_{0}^{\infty}dx e^{-x/\epsilon}
    \frac{1}{3x+2} \;.
\end{equation}
Applying the method of Ref. \citen{Ueoka14}, i.e. re-summing the series so that we obtain Eq. (\ref{eq:exponent_asymptotic}) in the limit of infinite
dimension, we get
\begin{equation}\label{eq:prevunitary}
   \nu  \approx \frac{1}{2} + \frac{1}{\epsilon^2} \int_{0}^{\infty}dx e^{-x/\epsilon}
    \frac{1}{5x+2} \;.
\end{equation}
We list the vales obtained for $d=3, 4, 5$ and $6$ in Table \ref{table:u_integer_d}.

\end{document}